\documentclass[12pt]{article}
\def\be{\begin{equation}}
\def\ee{\end{equation}}

\begin{document}
{\bf Multiparticle production: an old-fashioned view }\\

V.A. Schegelsky and M.G. Ryskin \\

\vspace*{0.5cm}

 Petersburg Nuclear Physics Institute, NRC ``Kurchatov Institute'',
Gatchina, St.~Petersburg, 188300, Russia \\

{\bf Abstract} We consider the dynamics of high energy multiparticle
 production and discuss how the space-time picture of inelastic interaction
may reveal itself in  identical particles Bose-Einstein correlations.\\

\section{Introduction}
The high energy interaction is usually described as the interaction of coloured
particles (quarks and gluons) mediated by gluon exchange. On the other hand due
to confinement we can observe the {\em colourless} meson or baryons only. That
is the same final amplitude of a few(colourless) particles production should be
described in terms of colourless objects (degrees of freeedom). Recall that in
initial state we deal with the proton or mesons or photon which has no colour.
In terms of QCD these incoming particles may be treated as some colour dipole
(or higher multipole) formed by constituent quarks (and gluons). The
interaction (say, after the gluon exchange) changes the colours of initial
constituents. The new state can be viewed as a few new dipoles (multipoles),
Each of them is again colourless but the mass of this new dipole is now large.
The simplest example is the quark-antiquark pair produced by the heavy photon
in $e^+e^-$ annihilation.
 Originally this pair was produced at very small distances but the momenta of
the quark and the antiquark are large and directed in opposite sides. So the
separation between the quarks (and the dipole moment) increases.
Correspondingly increases the strenght of colour electric field between these
quarks. When the strenght of colour field becomes sufficiently large the new
quark-antiquark pair is produced from the vacuum fluctuations. Since this pair
is produced from the vacuum it is colourless. This new pair breaks the colour
string and as a result we obtain two new dipoles\footnote{For simplicity we
talk here about the dipoles only.}, each of a smaller mass. The situation is
repeated in each new dipole, so that finally we observe a system of few dipoles
with the mass of the order of the ordinary hadron mass which should be
identified with the secondary hadrons.

Recall that this picture is based on the famous Swinger model - QED in 1+1 dimensions.

Starting with the energy $\sqrt s$ this way we get the final
 multiplicity $N\propto \ln s$ and not $N\propto \sqrt s$. In other words it is enough to beak
 the string $\ln s$ times to provide the low ($\sim 1$ GeV) mass of each dipole and a finite
 (say, $\sim 3-10$ GeV) energy of the pair of the nearest dipoles.\\

Such a picture for the first time was implemented in the LUND-string
model~\cite{BG}.
 It is the key-stone of the hadronization routing used in many general purpose Monte Carlo generators.\\

We have to emphasize that all the process of multiparticle production was described above in terms
of the colourless objects --  incoming dipoles, new colourless pairs produced from the vacuum
fluctuations and the secondary dipoles.
That is it can be reworded in terms of the colourless (hadron) degrees of freedom.
Indeed, many years ago V.N. Gribov discussed the possibility to treat the Pomeron
 as the high energy vacuum fluctuation which contains the $\ln s$ Feynman
partons including the "wee" partons with low rapidities in the wave function of a fast hadron.~\cite{VNG}.
 Then the multiparticle production from one or few 'cut' Pomerons might
be calculated with the help of the AGK cutting rules~\cite{AGK}.

Note that the picture predicts the plato in rapidity distribution of secondaries. Moreover, since QCD is the logarithmical theory we have to expect  a slow growth of the mean transverse momenta with the energy increasing due to a larger avaliable phase space. This is the origing of the fact that asymptotically  the slope of the BFKL Pomeron tragectory $\alpha'_{BFKL}\to 0$~\cite{BFKL}.\\

 The detailed  experimental study of multiparticle production became possible
 only after start of the first hadron collider at CERN. UA1
experiment ~\cite{ua1bec} has made  somewhat unexpected discovery: the radius
of a radiation source in $pp$ collisions  at centre-off-mass energy from 0.2 to
0.9 TeV depends on particles multiplicity but not on beam energy. This feature
quite naturally might be described in the frame of multiperiperal model
\cite{SKMR,RS}.
 \\

\section{The size measured by Bose-Einstein correlations}
 The size of the secondary hadrons source measured via the
Bose-Einstein correlation (BEC) of identical pions radiated by the same colour
string (or by the one cutted Pomeron) does not increases with energy. It is
determined by the internal structure of the string or by the mean transverse
momenta, $p_T$, of the particles which form the Pomeron\footnote{In the case of
the multiperipheral model for the Pomeron the size is driven by the $p_T$ of
the hadrons in the multiperipheral ladder.}. On the other hand the total cross
section and the interaction radius measured by the $t$-slope of elastic
amplitude increase with energy. This large interaction radius is generated by
 more complicated diagrams with the multi-Pomeron exchange or with a
few strings. That is in BEC we have to see at least two different radii - one
corresponding to the correlation of two pions emitted from the same
string/Pomeron and another one caused by the observation of two pions radiated
from two different strings/Pomerons. The second radius should be larger.

In the recent paper~\cite{TS} it was proposed to fit the data on BEC by the
formulae which contains two different scales in order to extract both the
radius corresponding to an individual string or Pomeron and the radius caused
by the separation between the different strings/Pomerons. Note that the
relative contribution of these two component may depend on different factors.
In particular, on the multiplisity in a given event and/or on the mean
transverse momentum of the pions in the pair. Indeed, in the case with a larger
number of cut Pomerons (or strings) we expect a larger multiplicity. That is in
a low multiplicity events we observe mainly the pairs radiated from the same
(one) Pomeron while at a high multiplicities we deal mainly with the two pions
emitted from the different strings/Pomerons.

On the other hand,since the inclusive cross section falls down steeply with the pion $p_T$  selecting the pairs
with a larger $k_T=(p_{T1}+p_{T2})/2$ already at $k_T\sim 1$ GeV we actually will get the pairs where both pions
are produced from the same minijet with jet $E_T\sim  few$ GeV.
 In such a case the BEC will gives the radius of the parent minijet.\\

It is amusing to mention that all features of the radiated Pomeron presented
above might be recognized  in Fig.3 of the CMS publication \cite{CMS}. Later
similar results have been published by the ATLAS \cite{ATLAS}. Recall that in
these papers the analysis was done using the fit with only {\em one} radius
which, depending on the multiplicity of a particular event and $k_T$ of the
pair, should be considered as some ``average'' between the small raduis of an
individual source and the source-source separation.  Nevertheless:
\begin{itemize}
\item
At small small multiplicity, one might see ONLY small "radiator", $\sim 1fm$.
\item
The size of this object depends on BEC pairs transverse momentum : the higher
this momentum the better the resolution of BEC-femtoscope.
\item
The size of this object is independent on beam energy.
\item
The correlation strength ($\lambda$) at the smallest multiplicity has a maximal
value ($\lambda\simeq 1$), giving no chance for other radiation sources.
\item
At high multiplicity, the size  of radiation zone is mainly  determined by the
pions radiated from different Pomerons, i.e. it depends on Pomerons spatial
distribution over the radiation region. To be more precise, observed radius of
the radiation zone is a "weighted" mean of the separations between the
%radii value  of  a large number of
Pomerons ( at multiplicity $\sim 100$ there are $\sim 10-15$ Pomerons) and much
larger values of the radius observed via BEC (with only one scale) in a large multiplicity events corresponds to the large distances between Pomerons. Under such conditions the
strength of correlations decreases with multiplicity ( "sharing" between
different "processes").
\end{itemize}

In terms of the quark-gluon degrees of freedom we may say that experiment favor
the scenario
  where the hadronization of the quark-gluon system/matter/liquid passes via the formation of a number
  of relatively small size colourless bubbles/drops which finally produces the hadrons.

\section{Conclusion}
The experiments indicates that secondaries are produced by some small size
sources distributed over a much larger (of order of the whole radiation
size ) domain. These sources may be considered as the individual Pomerons or as
the minijets or the colour strings between the jets which emit the 'spray' of
hadrons. Note that the value of small radius measured in a low multiplicity
events (where we observe the one cut Pomeron or the low $p_T$ colour tube/string
between the incoming constituents) is (up to experimental accuracy) the same as
that for the relatively large $k_T$ pairs (where we study the minijet
fragmentation).
 That is in both cases we deal with the hadronization of the {\em same}
colour string.\\

\section*{Acknowledgement}
MGR was supported by the RSCF grant 14-22-00281.


\thebibliography{}
\bibitem{BG} (Lund-string)
%  A Semiclassical Model for Quark Jet Fragmentation
  Bo Andersson, G. Gustafson, C. Peterson
 Z.Phys. {\bf C1} (1979) 105 \\
%Parton Fragmentation and String Dynamics
  Bo Andersson, G. Gustafson, G. Ingelman, T. Sjostrand
 Phys.Rept. {\bf 97} (1983) 31.

\bibitem{VNG} V.N.Gribov,
 The Theory of Complex Angular Momenta,Appendix A1,
 Cambridge University Press, 2003.

\bibitem{AGK} V. Abramovsky, V.N. Gribov and O.V. Kancheli, Sov. J. Nucl. Phys. {\bf 18} (1964) 308.

\bibitem{BFKL} L.N. Lipatov,
%  The Bare Pomeron in Quantum Chromodynamics
%  L.N. Lipatov (St. Petersburg, INP). Nov 1985. 32 pp.
 Sov.Phys.JETP {\bf 63} (1986) 904 [Zh.Eksp.Teor.Fiz. {\bf 90} (1986) 1536].

\bibitem{ua1bec}  UA1 Collaboration: C. Albajar et al., Phys. Lett. {\bf B226}, 410 (1989).

\bibitem{SKMR} %V.A. Schegelsky, A.D. Martin, M.G. Ryskin and %V.A. Khoze,
V.~A.~Schegelsky, A.~D.~Martin, M.~G.~Ryskin and V.~A.~Khoze,
 %``Pomeron universality from identical pion correlations at the LHC,''
 Phys.\ Lett.\ B {\bf 703}, 288 (2011)
%� doi:10.1016/j.physletb.2011.07.085
 [arXiv:1101.5520 [hep-ph]].

\bibitem{RS} % M.G. Ryskin,V.A. Schegelsky,
 M.~G.~Ryskin and V.~A.~Schegelsky,
 %``Gribov-Regge Pomeron and hadron structure phenomenology at high energy,''
Nuclear Physics B (Proc. Suppl.) {\bf 219} (2011) 10�16.


\bibitem{TS}
%  Two scales in Bose-Einstein correlations
  V.A. Khoze, A.D. Martin, M.G. Ryskin, V.A. Schegelsky,
 arXiv:1601.08081 [hep-ph]

\bibitem {CMS} CMS Collaboration, V. Khachatryan et al., JHEP {\bf 1105} (2011) 029.

\bibitem {ATLAS} ATLAS Collaboration, G. Aad et al., Eur. Phys. J. {\bf C75} (2015) 466.
\end{document}